\documentclass[12pt,preprint]{aastex}

\usepackage{graphics,graphicx}

\slugcomment{To appear in {\it The Astrophysical Journal}}

\shorttitle{X-ray Emission from [WC] PNs}
\shortauthors{Montez et al.}

\begin{document}


\title{X-ray Imaging of Planetary
  Nebulae with Wolf-Rayet-type Central Stars: 
  Detection of the Hot Bubble in NGC 40}


\author{Rodolfo Montez, Jr.\altaffilmark{1}}

\author{Joel H. Kastner\altaffilmark{2}}

\author{Orsola De Marco\altaffilmark{3}}

\and 

\author{Noam Soker\altaffilmark{4}}

\altaffiltext{1}{Department of Physics and Astronomy, University of Rochester, Rochester, NY 14627; rudy@pas.rochester.edu.} 
\altaffiltext{2}{Chester F. Carlson Center for Imaging Science, Rochester Institute of Technology, 54 Lomb Memorial Dr., Rochester, NY 14623; jhk@cis.rit.edu.} 
\altaffiltext{3}{Department of Astrophysics, American Museum of Natural History, Central Park West at 79th Street, New York, NY 10024; orsola@amnh.org.} 
\altaffiltext{4}{Department of Physics, Technion-Israel Institute of Technology, 32000 Haifa, Israel; soker@physics.technion.ac.il.} 




\begin{abstract}

We present the results of {\it Chandra} X-ray Observatory
(CXO) observations of the planetary nebulae (PNs) NGC 40 and
Hen 2-99.  Both PNs feature late-type Wolf-Rayet central
stars that are presently driving fast ($\sim1000$ km
s$^{-1}$), massive winds into denser, slow-moving ($\sim$10
km s$^{-1}$) material ejected during recently terminated
asymptotic giant branch (AGB) evolutionary phases. Hence,
these observations provide key tests of models of wind-wind
interactions in PNs. In NGC 40, we detect faint, diffuse
X-ray emission distributed within a partial annulus that
lies nested within a $\sim40''$ diameter ring of nebulosity
observed in optical and near-infrared images. Hen 2-99 is
undetected. The inferred X-ray temperature ($T_X \sim10^6$
K) and luminosity ($L_{\rm X} \sim2\times10^{30}$ ergs s$^{-1}$)
of NGC 40 are the lowest measured thus far for any PN
displaying diffuse X-ray emission. These results, combined with
the ring-like morphology of the X-ray emission from NGC 40,
suggest that its X-ray emission arises from a ``hot bubble''
that is highly evolved and
is generated by a shocked, quasi-spherical fast wind
from the central star, as opposed to AGB or post-AGB jet
activity. In constrast, the lack of detectable X-ray emission from Hen
2-99 suggests that this PN has yet to enter a phase of
strong wind-wind shocks.

\end{abstract}







\section{Introduction}

Planetary nebulae (PNs) are the descendents of intermediate
($1-8$ $M_\odot$) mass stars.  The central star of a PN will
evolve into a white dwarf with a mass between $\sim0.55$ and $\sim1.0$
$M_\odot$.  While on the upper asymptotic giant branch (AGB),
much of the star's initial mass is lost in a more or less
spherical outflow at rates of up to $10^{-4} M_\odot$ yr$^{-1}$ and at
speeds of $\sim10-20$ km s$^{-1}$.  As the star leaves the AGB,
the UV radiation from the emerging white dwarf (WD) ionizes
the red giant ejecta, producing the short-lived ($\sim3\times10^4$ yr)
PN. The star's surface escape velocity also
increases at this stage, and its wind speed may increase to
$\sim1000$ km s$^{-1}$ as its mass loss rate drops to $\sim10^{-8} M_\odot$
yr$^{-1}$. This fast wind is shocked as it collides with the
previously ejected slow wind. This process is expected to play a role in PN 
shaping. 

These same wind-wind shocks are energetic enough to generate
X-ray emission and, with the advent of the Chandra X-ray
Observatory and XMM-Newton, diffuse X-ray emission indeed
has been detected in several PNs
\citep{kastner00,kastner01,kastner02,kastner03,chu01,guerrero02,guerrero05}.
However, the degree to which wind-wind interactions contribute to
PN X-ray emission likely varies from object to object.  In
particular, some PNs display X-ray ``hot bubbles,''
as predicted by wind-wind interaction models (e.g., BD
+30$\circ$3639, \citet{kastner03}; NGC 7009,
\citet{guerrero02}), whereas the X-ray morphologies of other PNs indicate
that their X-ray emission is a direct consequence of jet activity
(e.g., He 3-1475, \citet{sahai03}; Mz 3, \citet{kastner03}).

Planetaries harboring relatively cool, WC-type Wolf-Rayet
central stars (typically designated [WCL] objects; e.g.,
\citet{leuen96})
represent important test cases for understanding the origin
and nature of X-ray emission from PNs. The
central stars of [WCL] PNs are
characterized by their overall spectral resemblance to
``bona-fide,'' massive WR stars, but as a group they extend
to both hotter and cooler effective temperatures \citep{gortyl00}. The
luminosities determined for the WC central stars of PNs,
given reasonable distance estimates, confirm that these
stars are much less luminous than the WC remnants of massive
stars and establish beyond doubt their post-AGB nature.

Key properties of the [WCL]s indicate they should offer prime
examples of strong wind-wind interactions in PNs.
The [WCL] central stars exhibit extreme hydrogen deficiency, 
similar to that of their likely progeny, the PG 1159-type pulsating
white dwarfs \citep{koesterke98}. By 
virtue of the high opacity characteristic of a C- and O-enriched, yet 
H-depleted gas mix \citep{dembar01}, the [WCL] central stars develop
powerful winds characterized by mass loss rates
$\sim10^{-6}$ $M_\odot$ yr$^{-1}$ \citep{leuen96}.
The central star wind velocities 
of [WCL]s range from $\sim200$ to $\sim1000$ km s$^{-1}$
and appear correlated with spectral type, 
wherein the coolest [WCL]s have the lowest wind speeds \citep{leuen96}.
These velocities are unusually large for such relatively cool 
central stars. The very large stellar wind momenta of [WC] PNs
appear to result in systematically larger nebular expansion
velocities among these objects, relative to PNs with H-rich
central stars \citep{gorsta95}. Furthermore, as a group, the [WCL] 
PNs are characterized by dense, blobby structures
and are rich in dust and molecular gas. These nebulae also 
tend to be relatively compact; most feature optically bright shells with
radii $<3000$ AU \citep{gortyl00} which,
given typical late AGB expansion velocities ($\sim15-30$ km s$^{-1}$), 
suggest they are quite young (dynamical ages $<1000$ yr). 

Taken together, the foregoing suggests that the H abundances
of [WCL]s sharply declined just at the end of their AGB
evolution \citep{herwig01}. Hence, these stars have acquired
large envelope opacities, resulting in strong winds --- much
stronger than in normal, H-rich central stars of
PNs. Furthermore, the strong winds have emerged very early
in post-AGB evolution, before the remnant, ejected AGB
envelope has had time to disperse into the interstellar
medium.  Thus, the [WCL] PNs should make ideal tests for the
theory of production of X-rays via interacting wind shocks,
as the very large wind momenta of [WCL] central stars and
the large, lingering masses of AGB ejecta in [WCL] PNs
should offer precisely the right conditions for production
of hot, post-shock gas at high emission
measure. Furthermore, the blobby and/or filamentary
structure of [WCL] PNs should favor heat conduction and
mixing of ``hot bubble'' and nebular gas \citep{sokastner03}.

Indeed, among the prototype [WCL] objects is BD $+30^\circ3639$, the
first well-established --- and brightest --- example of
diffuse X-ray emission from a PN
\citep{kreysing92,arnbor96,kastner00,kastner02,sokastner03}.
The central star within the second-brightest
diffuse X-ray PN, NGC 
6543, also has long been classified as Wolf-Rayet type
\citep{swings40}. Its wind 
speed and mass loss rate (1750 km s$^{-1}$ and $\sim10^{-7}$ $M_\odot$ 
yr$^{-1}$, respectively; \citet{perinotto89}) are quite large, explaining 
the star's broad emission line spectrum. However, most of the 
key optical emission lines characteristic of the central
stars of [WC] PNs \citep{crowther98} are not present in the  
spectrum of the central star of NGC 6543 (De Marco,
unpublished), and the star is not H 
deficient \citep{mendez88}. Therefore, a [WC] classification
is precluded. Nevertheless, NGC 6543 may represent a
transient stage in the evolution of [WC] PNs, as its central
star may belong to a potentially related, ``weak
emission line'' class \citep{acker96,pena01}. 

Additional X-ray observations of [WC] PNs should further
clarify the role of post-AGB fast winds in generating diffuse
X-ray emission within PNs. To this end, we used the {\it
Chandra} X-ray Observatory (CXO) to search for X-ray
emission from two well-studied [WC] PNs, NGC 40 (central
star spectral type [WC8], fast wind speed $v_f = 1000$ km
s$^{-1}$; \citet{leuen96}) and Hen 2-99 ([WC9], $v_f
= 900$ km s$^{-1}$). The low-resolution optical
morphologies (e.g.,
\citet{kaler89}), central star spectral types, fast wind 
speeds, and (large) central star mass loss rates
($\sim3\times10^{-6}$ $M_\odot$ yr$^{-1}$; \citet{leuen96}) of these two PNs
bear close resemblance to BD +30$^\circ$3639 ([WC9], $v_f =
700$ km s$^{-1}$, \citet{leuen96}; mass loss rate
$\sim10^{-6}$ $M_\odot$ yr$^{-1}$, \citet{sokastner03}). Unlike BD
+30$^\circ$3639 and NGC 6543 --- both 
of which have been the subject of intensive, multi-epoch
observing campaigns by the {\it Hubble Space Telescope}
(HST; \citet{balick04,li02}) --- neither NGC 40 nor Hen 2-99
has been the subject of deep HST imaging.  

\section{Observations \& Results}



The CXO observed Hen 2-99 and NGC 40,
with the back-illuminated CCD S3 of the Advanced CCD Imaging
Spectrometer (ACIS) as the focal plane instrument, on 2003
November 12 (ObsID 4480) and 2004 June 13 (ObsID 4481),
respectively.  Exposure times 
were 29 ks for Hen 2-99 and 20 ks for NGC 40.  ACIS has a
pixel size of $0.49''$ and the field of view of ACIS-S3 is
$\sim8'\times8'$. The CXO/ACIS combination is
sensitive over the energy range 0.3--10 keV.  The data
were subject to standard processing by the Chandra X-ray
Center pipeline software (CIAO, V. 2.3). We further applied
energy-dependent subpixel event position corrections
appropriate for back-illuminated CCD ACIS-S3 \citep{li03}.

\subsection{NGC 40}

\begin{figure}
\includegraphics[scale=0.75,angle=90]{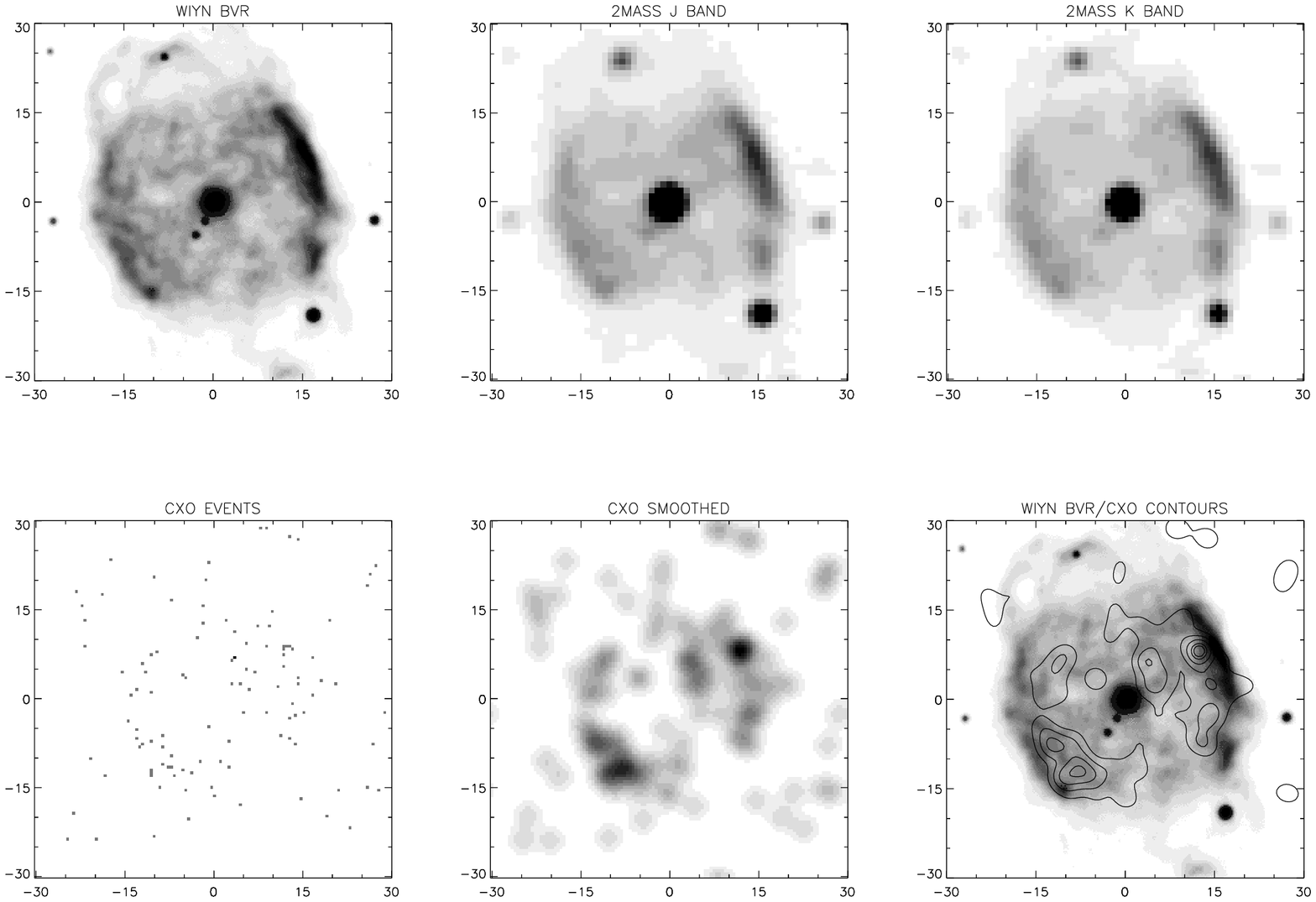} 
\caption{Optical, near-infrared, and X-ray images of NGC
  40. {\it Top row, left to right:} WIYN BVR composite 
  image (converted to greyscale) and 2MASS J- and K-band images.
  {\it Bottom row, left to right:} CXO X-ray and smoothed (see
  below) CXO X-ray images, and 
  contours of the smoothed CXO X-ray image overlaid on the
  WIYN BVR composite greyscale image.
  The images are $60''\times60''$ with N up and E to the
  left. The 2MASS images are displayed on a linear
  greyscale.  The smoothed CXO X-ray images in this Figure
  and in Figure 5 were
  obtained by convolving the original image
  with a PSF with FWHM $\sim4''$. The contours of the
  CXO X-ray image in the lower right panel are at $20\%$, 
  $40\%$, $60\%$, $80\%$, and $90\%$ of the image maximum of
  $1.6\times10^{-3}$ counts ks$^{-1}$ arcsec$^{-2}$ (0.3-1.0
  keV).}      
\label{fig:N40imgs}
\end{figure}

In Fig.~\ref{fig:N40imgs} we present WIYN\footnote{The WIYN
  Observatory is owned and operated by the WIYN Consortium,
  which consists of the University of Wisconsin, Indiana
  University, Yale University, and the National Optical
  Astronomy Observatories (NOAO). NOAO is operated by the
  Association of Universities for Research in Astronomy
  (AURA), Inc.\ under cooperative agreement with the
  National Science Foundation. See
  http://www.noao.edu/wiyn/wiynimages.html} optical (composite BVR) 
and 2MASS\footnote{This publication makes use of data
  products from the Two Micron All Sky Survey, which is a
  joint project of the University of Massachusetts and the
  Infrared Processing and Analysis Center/California
  Institute of Technology, funded by the National
  Aeronautics and Space Administration and the National
  Science Foundation. (http://www.ipac.caltech.edu/2mass/)}
  near-infrared (J- and K-band) 
images of NGC 40 along with the narrow-band (0.3--1 keV) Chandra/ACIS X-ray
image (raw and smoothed) of the same region. The morphology of the 
$\sim40''$ diameter nebula is similar in these three optical/near-infrared 
images: in each case NGC 40 appears as a limb-brightened
shell, with a bright rim that is interrupted by fainter
protrusions to the north-northeast and 
south-southwest. Deep optical images reveal jet-like
features in the vicinity of each protrusion, although there
is no kinematical evidence for collimated,
fast outflows in NGC 40 \citep{meaburn96}.

\begin{figure}
\centering
\includegraphics[scale=0.75,angle=0]{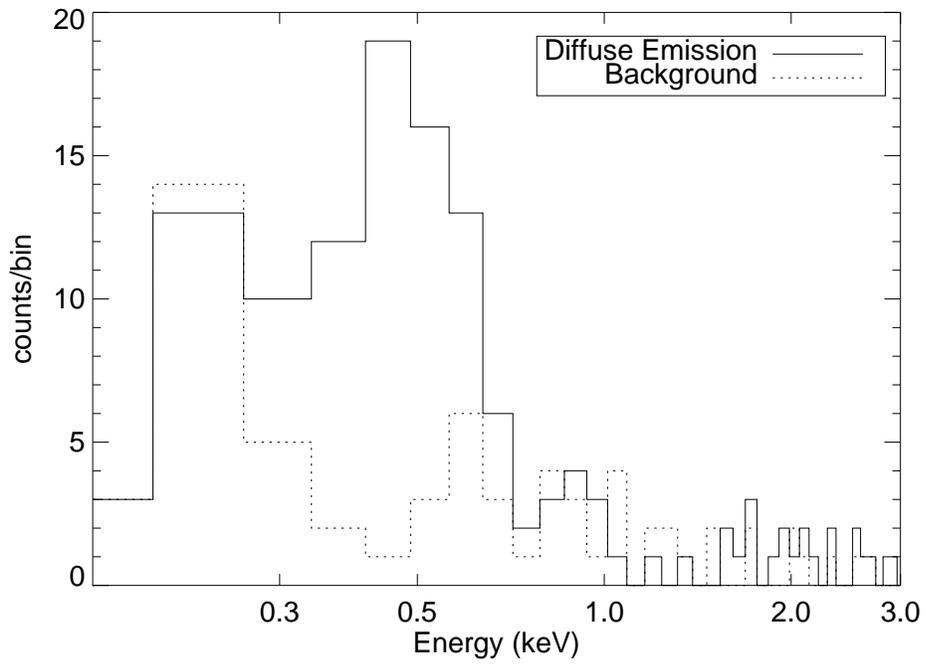} 
\caption{``Raw'' (counts vs.\ energy) X-ray spectrum
  of NGC 40 (solid histogram), extracted from the region encompassed by the
  optical/near-IR nebula (the raw background spectrum is displayed
  as a dotted-line histogram).}
\label{fig:N40rawspec}
\end{figure}

\begin{figure}
\centering
\includegraphics[scale=1.0,angle=0]{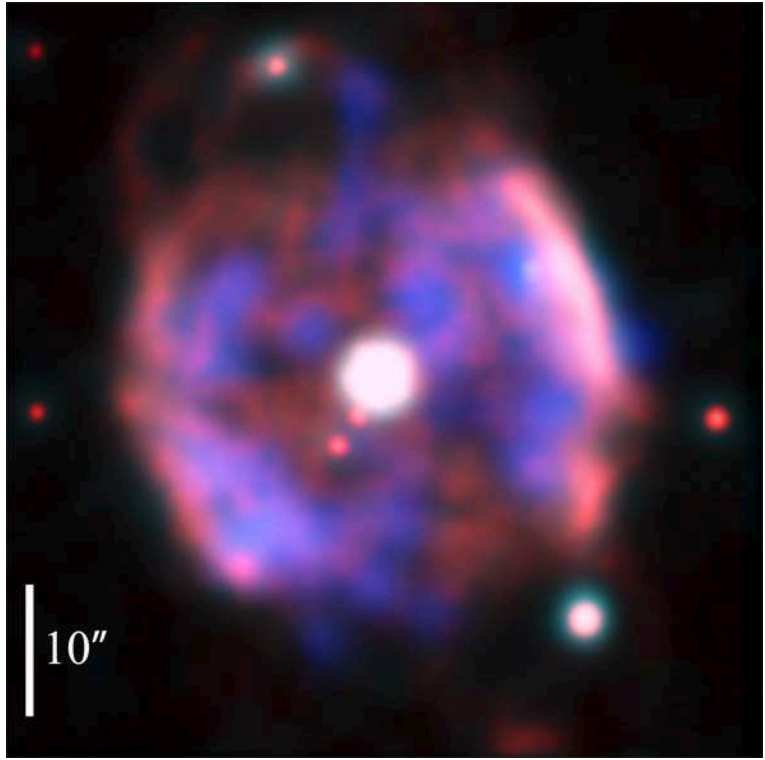} 
\caption{Color overlay of WIYN BVR composite (coded red),
  near-infrared 2MASS K-band (blue-green), and CXO X-ray images
  (blue) of NGC 40. The white regions of the nebula indicate
  a combination of strong optical (BVR) and K-band emission.} 
\label{fig:N40color}
\end{figure}

The detection of soft X-rays from NGC 40 is
apparent upon extraction of a spectrum
from a region of the Chandra/ACIS image encompassing the
optical/near-infrared nebula. This spectrum is displayed in
Fig.~\ref{fig:N40rawspec}. An image 
extracted over the soft 
(0.3-1.0 keV) energy range spanned by the detected photons
reveals that the X-ray emission  
arises from an annular region that lies nested just within,
and follows the overall surface brightness distribution of,
the bright, partial rim seen in the optical and near-infrared
(Figs.~\ref{fig:N40imgs}, \ref{fig:N40color}).    

\begin{figure}
\centering
\includegraphics[scale=0.6,angle=270]{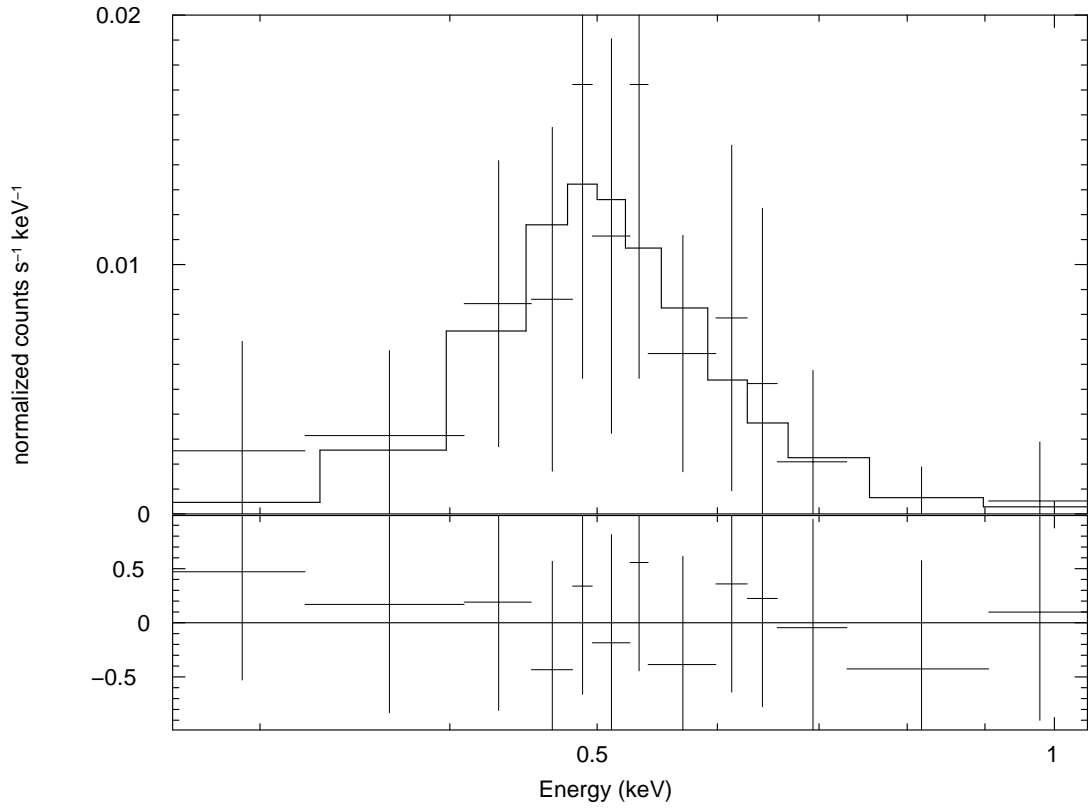} 
\caption{Top frame:
  Background-subtracted X-ray spectrum
  of NGC 40 (crosses), with best-fit thermal plasma model overlaid (solid
  histogram; see text). Bottom frame: residuals of the fit
  in units of $\sigma$.}     
\label{fig:N40spectrum}
\end{figure}

Refining the spectral extraction region to
an annulus with inner 
radius $4''$ and outer radius of $20''$ 
results in the background-subtracted spectrum displayed in
Fig.~\ref{fig:N40spectrum}.  The 
background-subtracted count rate of this diffuse emission
was $(2.8 \pm 0.9)\times10^{-3}$ counts s$^{-1}$, where the background
region was defined as a
circular region of radius $\sim40''$ lying  $\sim135''$
SE of NGC 40.  We used 
XSPEC (V.\ 12.2.0; \citet{arnaud96}) to fit the spectrum
with a Raymond-Smith thermal 
plasma emission model suffering intervening absorption.
Based on the inferred nebular color excess of $E(B-V) = 0.38$
\citep{pottasch77}, we fixed the absorbing column at $N_H =
2.2\times10^{21}$ cm$^{-2}$. The best fit indicates a 
plasma temperature of $T_x \sim 8.0\times10^5$ K, with an
uncertainty of $\sim20$\%.  The model fit is marginally improved
by the inclusion of a Gaussian component at an energy
$0.5\pm0.2$ keV, which is suggestive of the presence of
excess O {\sc vii} emission. The inclusion of this second
component would imply a somewhat higher plasma temperature
($T_x \sim 1.5\times10^6$ K). The modeling indicates an
unabsorbed X-ray flux $\sim1.3\times10^{-14}$ ergs cm$^{-2}$
s$^{-1}$ (the mean of the flux from plasma components
obtained by fits with and without the 0.5 keV Gaussian component),
corresponding to an intrinsic X-ray luminosity 
$L_{\rm X} \sim 1.5\times10^{30}\,(D/1.0 {\rm kpc})^2$ ergs
s$^{-1}$ where $D$ is the distance to NGC 40. The
estimate $D = 1.0$ kpc was determined by \citet{leuen96}.

\subsection{Hen 2-99}

\begin{figure}
\includegraphics[scale=0.75,angle=90]{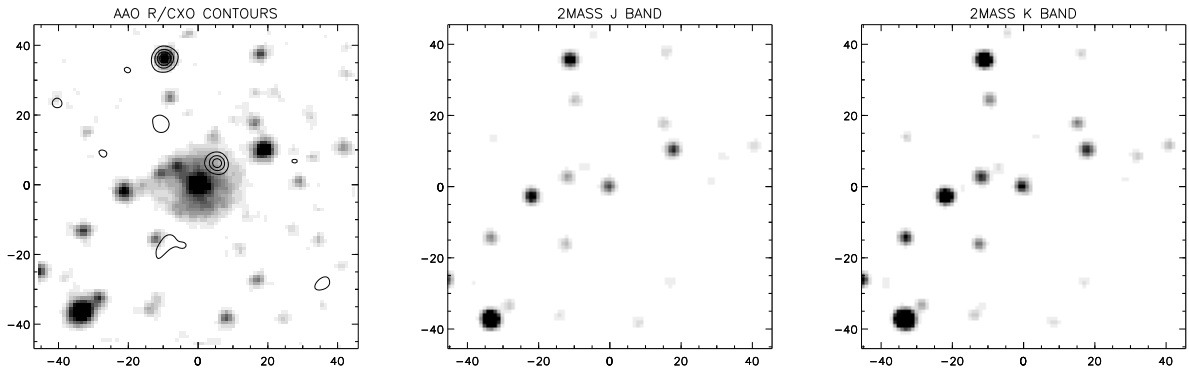} 
\caption{Optical, near-infrared, and X-ray images of the Hen
  2-99 region. {\it Left:} AAO Short R-band image, with smoothed
  CXO X-ray image overlaid as a contour map. {\it Center:}
  2MASS J-band image. {\it Right:} 2MASS K-band image.
  The images are $90''\times90''$ with N up
  and E to the left. The AAO and 2MASS images are presented as
  linear greyscale.  The contours in the left panel are at $20\%$,
  $40\%$, $60\%$, $80\%$, and $90\%$ of the X-ray image maximum of
  0.14 counts ks$^{-1}$ arcsec$^{-2}$ (0.3-10.0 keV).}
\label{fig:Hen2-99imgs}
\end{figure}

In Fig.~\ref{fig:Hen2-99imgs} we present Anglo-Australian
Observatory Digitized Sky Survey\footnote{http://archive.eso.org/dss/dss} 
Short R-band and 2MASS J- and K-band
images of Hen 2-99 
along with the broad-band (0.3--10 keV) Chandra/ACIS X-ray
image (raw and smoothed) of the same region. The object appears as a
$\sim25''$ diameter nebula in the R-band image, while only the
central star is detected in the near-infrared image. 
Point-like X-ray sources in the immediate vicinity of Hen
2-99 --- including one source that lies $\sim6''$ from the
central star and has no optical or near-infrared counterpart
--- have similar and relatively hard spectra, 
and are most likely background sources. 

We searched for diffuse emission from Hen 2-99 via the same
approach used for NGC 40, i.e., we
extracted a 0.3--10 keV spectrum from the 
region of the Chandra/ACIS image
encompassing the R-band emission from Hen 2-99. The spectral
extraction avoided
the point source that lies closest to the position of the
central star. Upon
subtraction of background emission within a region defined by a circle with radius $\sim10''$, the
resulting X-ray spectrum (not shown) reveals no
source photons.  

From the $1 \sigma$ uncertainty in the
background count rate ($2\times10^{-3}$
counts s$^{-1}$) we find a (3 $\sigma$) upper 
limit on the unabsorbed X-ray luminosity of 
$L_{\rm X} < 5\times10^{30}\times(D/2.5 {\rm kpc})^2$ 
ergs s$^{-1}$.  The
estimate $D = 2.5$ kpc was determined by \citet{leuen96}.
For this upper limit, we
(conservatively) assume an X-ray emission 
temperature at the upper end of the range determined thus far
for the diffuse emission from PNs ($T_x = 3\times10^6$ K),
and adopt $N_H = 2.9\times10^{21}$ cm$^{-2}$ 
based on a color excess of $E(B-V) = 0.5$  \citep{leuen96}.

\section{Discussion}

The shell-like shape of the X-ray emitting 
region in NGC 40, and the correspondence of this region to the bright
optical/near-IR rim in this nebula, indicates that the
bright rim has been generated by the shocked fast wind from 
the central star. The lack
of X-ray emission associated with the apparent
``blowouts'' (interruptions in the rim) that are aligned along an
axis running NNE-SSW in NGC 40 further
suggests that this wind is more or less
spherically symmetric, with a possible enhancement along the
equatorial plane of the system rather than along the polar
axis. We note that there is no kinematic evidence for the
presence of collimated jets in NGC 40, despite the
appearance of polar ``blowouts'' and jet-like features in
this nebula \citep{meaburn96,martin02}. 

The X-ray morphology of NGC 40 appears similar to (but is
better resolved than) those of BD +30$^\circ$3639
\citep{kastner00} and NGC 7009 \citep{guerrero02}. The
appearances of these
structures are fully consistent with the generation of ``hot
bubbles'' via wind-wind interactions \citep{akashi06}. These
X-ray morphologies stand in sharp contrast to the radially
directed X-ray structures within the young planetary nebulae NGC
7027 and Mz 3, however. These two PNs display X-ray evidence for
the action of collimated jets
\citep{kastner01,kastner02,kastner03,cox02} and, in both
cases, the X-ray emission appears intimately associated with
the ongoing structural changes apparent in the
optical/near-infrared nebulae.  Thus it appears that two
quite different processes --- jets on the one hand, and
quasi-spherical fast winds on the other --- may be
responsible for energetic shocks and, hence,
high-temperature plasma in PNs (see also \citet{guerrero05} and
\citet{akashi06}). It is clear, furthermore, that the
different processes responsible for X-ray emission are
closely tied to the different PN shapes that are seen in
optical emission lines.

The inferred X-ray temperature and luminosity of
NGC 40 ($T_X \sim10^6$ K and $L_{\rm X} \sim 1.5\times10^{30}$
ergs s$^{-1}$, respectively; \S 2.1) are the lowest measured
thus far for any PN
\citep{sokastner03,kastner03,guerrero05}, and the radii of 
the bright rim and X-ray emitting 
shell in NGC 40 are $\sim10$ times larger than these same
structures in BD +30$^\circ$3639. In light of the similarity
of the properties of the central stars of NGC 40 and BD
+30$^\circ$3639, the comparison between physical sizes and
X-ray emission properties indicates that the former PN might
best be considered as an evolved version of the latter.
This interpretation places constraints on [WC] PN evolutionary
schemes wherein the central stars evolve from late to early
[WC] spectral types
(see \citet{pena01} and references therein). Specifically, 
given the subtle differences 
between the central star spectral types ([WC9] and [WC8],
respectively) and wind properties of NGC 40 and BD
+30$^\circ$3639 \citep{leuen96}, the contrast in their X-ray and
optical/near-infrared properties indicates that
[WCL] stars and their associated PNs may evolve between
subclasses on timescales $<5000$ yr, i.e., a time period shorter
than the approximate dynamical age of the hot bubble in NGC
40 \citep{akashi06}. We stress, however, that the notion that [WC] PN
central stars evolve from late to early subtypes remains the
subject of debate \citep{pena01}. 

The upper limit on X-ray luminosity that we
obtain for Hen 2-99 ($L_{\rm X} < 5\times10^{30}$ ergs s$^{-1}$;
\S 2.2) constrains this nebula 
to be less X-ray luminous than all PNs thus far detected in
X-rays \citep{sokastner03,kastner03,guerrero05} with the
exception of NGC 40 (this paper).  The nondetection of Hen
2-99 by CXO therefore would suggest either that the fast
wind from the central ([WC9]) star has yet to collide with
the ionized ejecta seen in optical images, or that any
diffuse X-ray emission emanating from shocks produced by its
fast wind has now faded to a level similar to
or less than that detected in NGC 40. The former
interpretation would appear to be more consistent with the
lack of a clearly defined inner rim in Hen 2-99 in
near-infrared images (Fig.\ 5), if the presence of such
limb-brightened bubble structures in optical/near-infrared
images of PNs is interpreted as evidence of wind-wind
shocks. Additional X-ray observations of [WC] PNs with and
without well-formed optical/near-infrared bubbles are
required to test the hypothesis that the lack of detectable
X-ray emission from Hen 2-99 indicates that this PN has yet
to enter a phase of strong wind-wind shocks, despite the
very large wind momentum of its central star.



\acknowledgments
This research was supported by NASA through Chandra award
number GO4--5169X issued to Rochester Institute of
Technology by the Chandra X-ray Observatory Center, which is
operated by Smithsonian Astrophysical Observatory for and on
behalf of NASA under contract NAS8--03060. O.D. is grateful
to Janet Jeppson Asimov for financial support.

\facility{CXO(ACIS)}




\clearpage

\end{document}